\documentclass[12pt,a4paper]{article}
\usepackage{amssymb,amsmath,amscd,epsfig}
\title{Constraints on neutrino mixing angle $\theta_{13}$ and
Supernova neutrino fluxes from the LSD  neutrino signal from SN1987A }
\author{
 Oleg Lychkovskiy \thanks{e-mail:
lychkovskiy@mail.ru}\hspace*{2mm}$^{\rm a,b}$
\\ ${\rm ^a}$ {\small\it Institute for Theoretical and Experimental Physics}\\
{\small\it 117218, B.Cheremushkinskaya 25,
Moscow, Russia}\\
${\rm ^b}$ {\small\it Moscow Institute of Physics and Technology
}\\{\small\it 141700, 9, Institutskii per., Dolgoprudny, Moscow
Region, Russia}}

\date{}

\begin{document}
\maketitle

\begin{abstract}
Detection of 5 events by the Liquid Scintillation Detector (LSD)
on February, 23, 1987 was recently interpreted as a detection of
the electron neutrino flux from the first stage of the two-stage
Supernova collapse \cite{Imshennik}\cite{Gaponov}. We show that,
if neutrino mass hierarchy is normal, such interpretation excludes
values of neutrino mixing angle $\theta_{13}$ larger than $3\cdot
10^{-2}$, independently of the particular Supernova collapse
model. Also constraints on the original fluxes of neutrinos and
antineutrinos of different flavours are obtained.
\end{abstract}

{~~~~~~~~~~~~~~~~~~~~~~~~~~~~~~~~~~~~~~~~~~~~~~~~~~~~~~~~~~~~~~~~PACS-2006: 14.60.Pq; 26.50.+x.}\\
{Key words: {\it supernova neutrino; neutrino, mass spectrum; neutrino, mixing angle; \\SN1987A; neutrino; supernova.}}\\
\newline
Neutrino signal from the Supernova SN1987A, which explosion was
observed on February, 23, 1987, was detected by four experiments.
LSD experiment reported the detection of 5 events at the Unitary
Time 2:52UT, February 23, 1987 ~\cite{LSD1},\cite{LSD2}, while
KII, IMB and Baksan experiments reported the detection of 11, 8
and 5 events, correspondingly, at 7:35UT February 23,
1987~\cite{KII},\cite{IMB},\cite{Baksan}. We shall call 2:52UT the
LSD time, and 7:35UT -- the IMB time. KII, IMB and Baksan did not
see any considerable cluster of events at the LSD time, while LSD
did not see such a cluster at the IMB time. A description of the
combined time sequence of all detected events may be found in
\cite{Imshennik}. For a long time it was believed that all four
detectors are sensitive almost only to the electron antineutrinos,
KII and IMB being several times more sensitive than LSD. Thus
absence of a signal in KII and IMB at the LSD time remained an
unsolved puzzle. Recently the following solution was proposed
\cite{Imshennik},\cite{Gaponov}: LSD included a large amount of Fe
and thus was sensitive to the energetic electron neutrinos
($\varepsilon_{\nu}\sim 40$MeV) through the reaction
\begin{equation}\label{nu-Fe reaction}
\nu_e+\mathrm{Fe}\longrightarrow \mathrm{Co}^*+e^-.
\end{equation}
It was a unique LSD feature in comparison with three other
detectors. According to \cite{Imshennik}, \cite{Gaponov} the LSD
neutrino signal is interpreted as the detection of 5 electron
neutrinos of energies of about $40$MeV. This corresponds to the
total electron neutrino flux at the Earth
\begin{equation}\label{eperimental result 1}
F_{\nu_e}\sim 10^{10} \mathrm {cm}^{-2}.
\end{equation}
This flux was too small to produce a considerable cluster of
events in KII, IMB and Baksan, as their sensitivity to the
electron neutrinos was lower.

Total flux of electron antineutrinos at the Earth at the LSD time
had to be much smaller,
\begin{equation}\label{eperimental result 2}
F_{\bar{\nu}_e}\lesssim 10^{9} \mathrm {cm}^{-2},
\end{equation}
otherwise it would be detected by KII \cite{KII} and IMB
\cite{IMB}. The crucial for the subsequent argumentation fact is
that
\begin{equation}\label{main experimental
bound} F_{\bar{\nu}_e}\lesssim 10^{-1} F_{\nu_e}.
\end{equation}

As for the absence of the event cluster in LSD at the IMB
time, it is explained by two factors:\\
1. small LSD electron antineutrino sensitivity (compared to
other detector sensitivities),which prevented LSD from the electron antineutrino detection; \\
2. softness of the spectrum of electron neutrinos at the IMB time,
which, due to the high energy threshold of the reaction
(\ref{nu-Fe reaction}), prevented LSD from the electron neutrino
detection.

The existence of two neutrino signals from SN1987A, separated by
more than four hours, implies that the Supernova core collapse
proceeded in two stages. One of the models of such a collapse,
aimed at the explanation of the LSD signal, was proposed by
Imshennik and Ryazhskaya in \cite{Imshennik}. According to it, at
the first stage of the collapse almost exclusively electron
neutrinos were produced, electron neutrino luminosity being
$L_{\nu_e}\sim 10^{53}$ erg, which is sufficient to produce
electron neutrino flux (\ref{eperimental result 1}). However,
Imshennik and Ryazhskaya did not take into account neutrino
flavour conversion inside the Supernova progenitor star. Lunardini
and Smirnov pointed out (see note added to \cite{Smirnov 0}) that
taking such conversion into account breaks the agreement between
the collapse model \cite{Imshennik} and the LSD data. Namely, if
the neutrino mass hierarchy is normal ($m_3>m_1,m_2$) and neutrino
mixing angle $\theta_{13}$ is larger than $\sim 10^{-2}$, then all
neutrinos produced as electron ones are converted to the mixture
of $\mu$- and $\tau$-neutrinos and thus can not be detected by
LSD. If the neutrino mass hierarchy is inverted or $\theta_{13}\ll
10^{-2}$, then the electron neutrino flux at the Earth is
suppressed by the factor $\sin^2\theta_{12}\simeq 0.3$ and thus
the original luminosity $L_{\nu_e}$ three times larger than in
\cite{Imshennik} is required to explain the LSD signal. It should
be emphasized, that Lunardini and Smirnov regard their result as
an argument against the model by Imshennik and Ryazhskaya
\cite{Imshennik} rather than a constraint on the mixing angle
$\theta_{13}$ and electron neutrino luminosity $L_{\nu_e}$.

In this note we show, that if we admit the interpretation of the
LSD neutrino signal proposed in \cite{Imshennik},\cite{Gaponov}
and take into account the absence of the neutrino signal in KII
and IMB at the LSD time, we may constrain neutrino mixing angle
$\theta_{13}$ in the case of normal neutrino mass hierarchy
independently of the particular Supernova collapse model. Also
constraints on the original neutrino fluxes (i.e. fluxes, produced
in a Supernova core) may be obtained. Constraints on the mixing
angle $\theta_{13}$ and original electron neutrino flux coincide
with the results of \cite{Smirnov 0}.

We make the following assumptions regarding original fluxes :
\begin{equation}\label{original fluxes}
F^0_x \equiv
F^0_{\nu_{\mu}}=F^0_{\nu_{\tau}}=F^0_{{\bar{\nu}_{\mu}}}=F^0_{{\bar{\nu}_{\mu}}}\leqslant
F^0_{\bar{\nu}_e},F^0_{\nu_e}.
\end{equation}
This assumptions are natural (see, for example, \cite{SN
simulation}). They follow from two facts. First, in Supernova
$\mu$- and $\tau$-neutrinos are always produced in pairs with
their antiparticles. Second, there are more electron
(anti-)neutrino production channels than non-electron ones:
$$e^++n\rightarrow\bar{\nu}_e+p,$$
 $$e^-+p\rightarrow\nu_e+n,$$
 $$A+A'\rightarrow A+A'+\nu+\bar{\nu}$$
 \begin{equation}\label{nu creatiion}
 e^++e^-\rightarrow\nu +\bar{\nu},
 \end{equation}
where $A,A'$ are nuclei, and $\nu$ stands for a neutrino of
arbitrary flavour .

Due to the matter effect neutrino flux of a given flavour is
modified in the outer layers of the Supernova progenitor star. As
detectors are sensitive almost only to electron neutrinos or
antineutrnos, we are interested in the electron neutrino and
antineutrino fluxes at the Earth $F_{\nu_e}$ and
$F_{\bar{\nu}_e}$. They are connected to the original fluxes as
follows \footnote{We omit obvious factor ${R^2_{ns}}/{R^2}$ in all
expressions for the fluxes at the Earth $F_{\nu_e}$ and
$F_{\bar{\nu}_e}$ (eqs. (\ref{final fluxes}), (\ref{inequality}),
(\ref{NL})--(\ref{NL once more time})). Here $R$ is the distance
between the Supernova and the Earth, $R_{ns}$ is the neutrino
sphere radius. In the spherically symmetrical case original
luminosity of neutrinos and antineutrinos of a given flavour may
be obtained from the corresponding original flux through $L=4\pi
R_{ns}^2 \varepsilon F,$ where  $\varepsilon$ is a mean
(anti-)neutrino energy.}
\cite{Smirnov}
:
$$F_{\nu_e}=pF^0_{\nu_e} + (1-p)F^0_x,$$
\begin{equation}\label{final fluxes}
F_{\bar{\nu}_e}=\bar{p}F^0_{\bar{\nu}_e}+(1-\bar{p})F^0_x,
\end{equation}
$p$ and $\bar{p}$ being the electron neutrino and antineutrino
survival probabilities. Note that from (\ref{original fluxes}) and
(\ref{final fluxes}) one may get
\begin{equation}\label{inequality}
F^0_x\leqslant F_{\nu_e}\leqslant F^0_{\nu_e},~~~~F^0_x\leqslant
F_{\bar{\nu}_e}\leqslant F^0_{\bar{\nu}_e}.
\end{equation}

Probabilities $p$ and $\bar{p}$ depend on the neutrino mass
hierarchy and on the value of the mixing angle $\theta_{13}$. We
distinguish three cases: {\bf NL} ({\bf N}ormal hierarchy, {\bf
L}arge angles), {\bf IL} ({\bf I}nverted hierarchy, {\bf L}arge
angles), {\bf AS} ({\bf A}ny hierarchy, {\bf S}mall angles). Here
large angles stand for $\theta_{13}\gtrsim 3\cdot 10^{-2}$, while
small angles stand for $\theta_{13}\lesssim 3\cdot
10^{-3}$. Remind that current experimental limit is $\theta_{13}<0.17$ ~\cite{CHOOZ}.  
 Survival probabilities are given by (see \cite{Smirnov}, \cite{Takahashi})
$$p=\sin^2 \theta_{13}\approx0,~~~\bar{p}= \cos^2 \theta_{12},~~~\mathbf{NL}$$
$$p= \sin^2 \theta_{12},~~~\bar{p}= \sin^2 \theta_{13}\approx0,~~~\mathbf{IL}$$
\begin{equation}\label{survival probabilities}
p= \sin^2 \theta_{12},~~~\bar{p}= \cos^2
\theta_{12},~~~\mathbf{AS}
\end{equation}
Let us consider this cases separately.
\\ {\bf{NL}}. From (\ref{final fluxes}) and (\ref{inequality}) we get

\begin{equation}\label{NL}
F_{\nu_e}\simeq F^0_x,~~~~F_{\bar{\nu}_e}\geqslant F^0_x.
\end{equation}
Consequently, $F_{\nu_e}\lesssim F_{\bar{\nu}_e}$, which is in
contradiction with experimental result (\ref{main experimental
bound}). Thus this case is excluded.\newline
\\{\bf{IL}}. From (\ref{final fluxes}) we get
\begin{equation}\label{IL1}
F_{\nu_e}= \sin^2 \theta_{12}F^0_{\nu_e} + \cos^2
\theta_{12}F^0_x,~~~~F^0_x \simeq F_{\bar{\nu}_e} ,
\end{equation}
thus, taking (\ref{main experimental bound}) into account,
\begin{equation}\label{IL2}
F^0_{\nu_e}= \frac{1}{\sin^2 \theta_{12}}(F_{\nu_e}-\cos^2
\theta_{12}F_{\bar{\nu}_e})\simeq 3F_{\nu_e}.
\end{equation}
Here we used $\sin^2 \theta_{12}\simeq 0.3$\newline
\\{\bf{AS}}. From (\ref{final fluxes}) we get
$$F_{\nu_e}= \sin^2 \theta_{12}F^0_{\nu_e} + \cos^2
\theta_{12}F^0_x$$
\begin{equation}\label{AS}
F_{\bar{\nu}_e}= \cos^2 \theta_{12}F^0_{\bar{\nu}_e}+\sin^2
\theta_{12}F^0_x,
\end{equation}
and, consequently,
$$F^0_x \leqslant F_{\bar{\nu}_e}
$$
$$F_{\bar{\nu}_e}\leqslant F^0_{\bar{\nu}_e}\leqslant\frac{1}{\cos^2
\theta_{12}}F_{\bar{\nu}_e}\simeq 1.4 F_{\bar{\nu}_e}$$
\begin{equation}\label{AS2}
F^0_{\nu_e}= \frac{1}{\sin^2 \theta_{12}}(F_{\nu_e}-\cos^2
\theta_{12}F^0_x)\simeq 3F_{\nu_e}.
\end{equation}

Note, that we did not take into account modifications of the
fluxes inside the Earth due to the matter effect, in spite of the
fact that neutrinos from SN1987A crossed the Earth on their way to
each of the detectors. A detailed study of the Earth matter
effects was performed in \cite{Smirnov},\cite{Smirnov 2}. From
this papers it follows that such effects provide a correction to
the fluxes of interest not greater than 20\% (for neutrino energies about 40 Mev).
 Thus  equalities and inequalities (\ref{IL1}-\ref{AS2}) are valid
with 20\% precision, which is better than statistical
uncertainties of determination of fluxes $F_{\nu_e}$ and
$F_{\bar{\nu}_e}$. As for eq.(\ref{NL}), it is not influenced by
the Earth matter effect at all. Indeed, in the {\bf NL} case the
correction to the electron neutrino flux $F_{\nu_e}$ is absent,
and corrected electron antineutrino flux reads (see
\cite{Smirnov},\cite{Smirnov 2})
\begin{equation}\label{Earth matter effect}
F_{\bar{\nu}_e}= (\cos^2\theta_{12}-f_{reg})
F^0_{\bar{\nu}_e}+(\sin^2\theta_{12}+f_{reg}) F^0_x,
\end{equation}
where $f_{reg}$ is a correction due to the Earth matter effect.
Independently of value of $f_{reg}$, this equation along with
inequality (\ref{original fluxes}) leads to
\begin{equation}\label{NL once more time}
 F_{\bar{\nu}_e}\geqslant F^0_x,
\end{equation}
which is exactly what we see in (\ref{NL}). Thus, accounting for
the Earth matter effect does not influence our conclusion that
{\bf NL} case is excluded.

Let us summarize the results. If one interprets 5 LSD events on
February, 23, 1987 as a detection of the electron neutrino flux
from the first stage of the two-stage Supernova collapse, one
comes to the following conclusions concerning neutrino properties
and original (anti-)
neutrino fluxes at the first stage of the explosion. \\
1. If the neutrino mass hierarchy is normal, than the case of
$\theta_{13}\gtrsim 3\cdot 10^{-2}$ is excluded.
\\2. Original $\nu_{\mu},~\bar{\nu}_{\mu},~\nu_{\tau}$ and $\bar{\nu}_{\tau}$ fluxes are less or equal
to the electron antineutrino flux at the Earth up to the obvious
factor ${R^2}/{R^2_{ns}}$, see eqs.(\ref{IL1}),(\ref{AS2}).
\\3. Original electron neutrino flux is three times larger than electron neutrino flux
at the Earth (up to the factor ${R^2}/{R^2_{ns}}$), see
eqs.(\ref{IL2}),(\ref{AS2}).
\\4. In the case of  $\theta_{13}\lesssim 3\cdot 10^{-3}$ original electron
antineutrino flux is of order of the electron antineutrino flux at
the Earth (up to the factor ${R^2}/{R^2_{ns}}$), see
eq.(\ref{AS2}).

The first conclusion is of particular interest, especially
regarding that if the neutrino mass hierarchy is inverted, large
values of the mixing angle, $\theta_{13}\gtrsim 3\cdot 10^{-2}$,
are disfavored by the combined KII and IMB data on the second
neutrino signal from SN1987A \cite{Smirnov 3}. Accordingly, the
total bulk of data on the SN1987A neutrino burst disfavors
$\theta_{13}$ larger than $3\cdot 10^{-2}$ whatever the mass
hierarchy is.

The last three conclusions constrain any model of the SN1987A
two-stage collapse. In particular, the second and the fourth
conclusions are consistent with the model of the first stage of
the SN1987A explosion by Imshennik and
Ryazhskaya \cite{Imshennik}, while the third one contradicts it.\\
\newline
{\bf\Large Acknowledgements}\\
 The Author wishes to thank A.A. Mamonov, V.S. Imshennik and L.B.
 Okun for valuable discussions and attention to this work. The
 work was financially supported by Dynasty Foundation and RF
 President grant NSh-5603.2006.2.



\begin{thebibliography}{99}

\bibitem{Imshennik}
  V.~S.~Imshennik and O.~G.~Ryazhskaya,
  Astron.\ Lett.\  {\bf 30}, 14 (2004)
  [arXiv:astro-ph/0401613].

\bibitem{Gaponov}
  Y.~V.~Gaponov, O.~G.~Ryazhskaya and S.~V.~Semenov,
  Phys.\ Atom.\ Nucl.\  {\bf 67}, 1969 (2004).

\bibitem{LSD1}
  V.~L.~Dadykin {\it et al.},
  JETP Lett.\  {\bf 45}, 593 (1987)
  [Pisma Zh.\ Eksp.\ Teor.\ Fiz.\

\bibitem{LSD2}
  M.~Aglietta {\it et al.},
  Europhys.\ Lett.\  {\bf 3}, 1315 (1987).

\bibitem{KII}
  K.~Hirata {\it et al.}  [KAMIOKANDE-II Collaboration],
  Phys.\ Rev.\ Lett.\  {\bf 58}, 1490 (1987).

\bibitem{IMB}
  R.~M.~Bionta {\it et al.},
  Phys.\ Rev.\ Lett.\  {\bf 58}, 1494 (1987).


\bibitem{Baksan}
  E.~N.~Alekseev, L.~N.~Alekseeva, V.~I.~Volchenko and I.~V.~Krivosheina,
  JETP Lett.\  {\bf 45}, 589 (1987)
  [Pisma Zh.\ Eksp.\ Teor.\ Fiz.\  {\bf 45}, 461 (1987)].

\bibitem{Smirnov 0}
  C.~Lunardini and A.~Y.~Smirnov,
  Astropart.\ Phys.\  {\bf 21}, 703 (2004)
  [arXiv:hep-ph/0402128].

\bibitem{SN simulation}
  M.~Liebendoerfer, O.~E.~B.~Messer, A.~Mezzacappa, S.~W.~Bruenn, C.~Y.~Cardall and F.~K.~Thielemann,
  Astrophys.\ J.\ Suppl.\  {\bf 150}, 263 (2004)
  [arXiv:astro-ph/0207036].



\bibitem{Smirnov}
  A.~S.~Dighe and A.~Y.~Smirnov,
  Phys.\ Rev.\ D {\bf 62}, 033007 (2000)
  [arXiv:hep-ph/9907423].

\bibitem{CHOOZ}
  M.~Apollonio {\it et al.},
  Eur.\ Phys.\ J.\ C {\bf 27}, 331 (2003)
  [arXiv:hep-ex/0301017].

\bibitem{Takahashi}
 K.~Takahashi and K.~Sato,
 Prog.\ Theor.\ Phys.\  {\bf 109}, 919 (2003)
 [arXiv:hep-ph/0205070].

\bibitem{Smirnov 2}
  P.~C.~de Holanda, W.~Liao and A.~Y.~Smirnov,
  Nucl.\ Phys.\ B {\bf 702}, 307 (2004)
  [arXiv:hep-ph/0404042].

\bibitem{Smirnov 3}
  C.~Lunardini and A.~Y.~Smirnov,
  Phys.\ Rev.\ D {\bf 63}, 073009 (2001)
  [arXiv:hep-ph/0009356].


\end{thebibliography}
\end{document}